# TOWARDS BALLISTIC TRANSPORT IN GRAPHENE

Xu Du, Ivan Skachko and Eva Y. Andrei

*Department of Physics and Astronomy, Rutgers University, Piscataway, New Jersey 08854*

Graphene is a fascinating material for exploring fundamental science questions as well as a potential building block for novel electronic applications. In order to realize the full potential of this material the fabrication techniques of graphene devices, still in their infancy, need to be refined to better isolate the graphene layer from the environment. We present results from a study on the influence of extrinsic factors on the quality of graphene devices including material defects, lithography, doping by metallic leads and the substrate. The main finding is that trapped Coulomb scatterers associated with the substrate are the primary factor reducing the quality of graphene devices. A fabrication scheme is proposed to produce high quality graphene devices dependably and reproducibly. In these devices, the transport properties approach theoretical predictions of ballistic transport.

## 1. Introduction

The discovery of techniques to isolate and study graphene, a one-atom thick layer of crystalline carbon[1-3], has stimulated a massive effort to understand its electronic properties[4,5]. As a Dirac fermion system with linear energy dispersion, electron-hole symmetry and internal degree of freedom (pseudo spin), graphene promises intriguing physical properties such as electronic negative index of refraction[6], specular Andreev reflections in graphene-superconductor junctions[7,8], evanescent transport[9], anomalous phonon softening[10], etc. As an electronic material, graphene exhibits many desirable properties such as high mobility, thin body, low carrier density (tunable by electric field gating), and compatibility with the top-down fabrication scheme.

Despite the high expectations for ideal graphene devices, the commonly used fabrication techniques yield samples with large uncontrollable variations and middling quality as characterized by mobility, gate control, and minimum conductivity. It is widely believed that this lower than expected performance is not intrinsic, but rather due to extrinsic factors such as material quality and ambient environment[11,12]. But it is not clear how the different factors contribute to the deterioration of the transport properties in graphene and how to asses their relative importance. Nor is there a scheme in place to produce high quality graphene devices reliably and reproducibly. Here we report results of a study that gauged the impact of various extrinsic factors on the quality of graphene devices. The basic strategy was to isolate the influence of each of the extrinsic factor on the transport properties of a freshly prepared graphene device by deliberately magnifying the corresponding factor to dominate all others. This method was used to study material defects, lithographic residues, invasiveness of metallic contacts, substrate. We find that, while all factors contribute to the degradation of the graphene device, trapped charges in the substrates are the main contributor. Based on our observations, a scheme is proposed for future fabrication of ultrahigh quality graphene devices.

## 2. Experiments and Results

Graphene is deposited using the method of mechanical exfoliation [1]. Prior to graphene deposition, the Si/SiO$_2$ substrates are baked in forming gas (Ar/H$_2$) at 200C for one hour to remove water and organic residue. A thin foil of highly oriented pyrolytic graphite (neutron detector quality) is peeled from the bulk material using scotch tape and transferred onto the Si/SiO$_2$ substrate. Pressure is then applied onto the graphite foil using compressed high purity nitrogen gas through a stainless steel needle, for ~5 seconds. The foil is then removed from the substrate and the substrate is carefully checked under an optical microscope for candidates of single layer graphene. This process is repeated until a few graphene flakes are identified. AFM inspection is then used to confirm the single graphene layer followed immediately by coating with PMMA resist. The electrical contacts and leads are then fabricated with standard e-beam lithography techniques. To remove the organic and water residues, the samples are baked in forming gas (Ar/H$_2$) at 200 C for 1 hour.

To test the impact of the material quality on transport properties of the graphene devices, we created defects in the graphene films with DC hydrogen glow discharge. For this purpose, pre-characterized graphene devices were placed ~5cm away from a circular electrode biased at 500V DC potential generating a glow discharge at a hydrogen pressure of 150 mTorr. The damage produced by 10 seconds exposure to the glow discharge is then characterized by comparing the gate voltage dependence of the sample resistance before and after irradiation. No noticeable change in the shape of the flake could be detected by optical microscopy or through SEM. We therefore attribute any changes in the transport properties to microscopic (atomic) defects from the irradiation. Figure 1a shows a typical result of the glow discharge tests. We note that the irradiation causes a significant shift of the Dirac point indicative of hole-doping, presumably due to the trapping of the positively charged hydrogen ions. The large increase of sample resistance throughout the whole range of gate voltage suggests a large increase in the density of scattering centers. Since the samples are in the diffusive regime, we were able to separate the Coulomb scattering contribution to the resistance (which yields gate-independent mobility) from the short-range scattering contribution (gate-independent resistivity), as shown in Figure 1b. Here the Coulomb scattering mobility is estimated as $\mu_C = 1/(\rho - \rho_S)ne$ where $n \sim 7.4 \times 10^{10} |V_g - V_D| \, cm^{-2}$ is the carrier density, $V_D$ the gate voltage at maximum resistivity. and $\rho_s$ is the short range scattering resistivity which is chosen so that the Coulomb scattering mobility $\mu_C$ is roughly independent of $V_g$ outside the e-h puddle regime[2, 13]. It is clear that irradiation with the glow discharge introduces both Coulomb and short-range scatterers: for the sample in Figure 1, the short-range resistivity increases from 350 to 1500 , and the Coulomb scattering mobility decreased from 6000 cm$^2$/Vs to 600 cm$^2$/Vs.

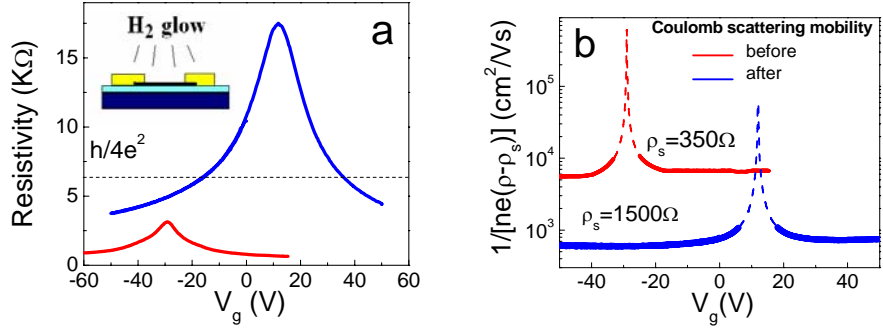

**Fig. 1. a.** Gate voltage dependence of resistivity before (red) and after (blue) the H2 glow discharge test. Inset. Scheme of the glow discharge treatment setup. **b.** Gate dependence of the estimated Coulomb scattering mobility before and after the glow discharge test. The dashed lines indicate e-h puddle regime in which the mobility cannot be estimated.

For graphene deposited on $SiO_2$ substrates, the typical observed maximum resistivity is $\rho_{max} \leq h/4e^2$. This is also the case for the sample shown here before the glow discharge treatment. After the irradiation, however the maximum resistivity became significantly larger than $h/4e^2$. Although the glow discharge caused damage to the material to an extent that is unlikely to appear in the commercially available high quality graphite, the test here suggests that the maximum resistivity in the graphene is not universal, but depends on material quality. In the extreme case of graphene with high defect density, the resistivity will consistently exceed that of high quality graphene throughout the entire doping range including the Dirac point.

Next, we discuss the impact of the lithographic process. Again, we start with pre-measured freshly made graphene devices. Subsequently the graphene is subjected to two lithographical processes: e-beam resist (PMMA) coating and Acetone stripping. Following every step of the treatment, the gate dependence of the resistivity is measured and compared to the pre-treatment measurement as illustrated in Figure 2a. We note that the PMMA coat shifts the Dirac point to negative gate voltage, indicating electron doping. Further analysis of the $\rho(V_g)$ curve suggests that the decrease in mobility associated with the PMMA coating is rather small (~20%) and can be attributed to Coulomb scattering. After the PMMA is removed with Acetone, the position of the Dirac point almost recovers to zero gate voltage and the entire $\rho(V_g)$ curve regains its pre-treatment shape with a slightly reduced mobility. These tests suggest that standard lithographic procedures do not significantly reduce device quality.

We next discuss the effect of contact leads. We focused on short aluminum-graphene-aluminum junctions with voltage leads that run across the sample width - 2-lead geometry, measuring the $\rho(V_g)$ curves as a function of lead separation. For lead separation shorter than 1μm, we observe clear particle-hole asymmetry in the $\rho(V_g)$ curves. For the shortest junctions, $L=$ 300 nm, the asymmetry is most pronounced we observe a hump forming on the hole branch. The asymmetry is not observed in samples with Hall bar lead geometry where the voltage leads do not run across the sample, or in samples with long channel length (L>>1μm). This suggests that the asymmetry is not intrinsic but rather that it is associated with the invasiveness of the contact leads. A possible explanation is that the asymmetry arises from doping by the metallic leads resulting in a locally altered Fermi energy relative to the Dirac point. This scenario can be modeled by considering the graphene channel as consisting of 3 sections: 2 doped sections near the contacts flanking one undoped graphene section. For long junctions where the center section is much longer than the doped sections, the undoped part of the channel dominates the total resistance, and the measured 2-terminal $\rho(V_g)$ dependence is roughly symmetric. As the junction becomes shorter, the contributions from the contact-doped sections become more important. The different positions of the Dirac point in the lead segments is different and in the undoped segment will thus cause an asymmetry in the $\rho(V_g)$ curve arising from the 2 Dirac points. For the central segment the Dirac point is rather sharp and sits at $V_g \sim 0$, while close to the leads it is broader (due to the variation of the Dirac point energy as a function of the distance from the leads) and its position is shifted towards large negative values of $V_g$ (electron doped).

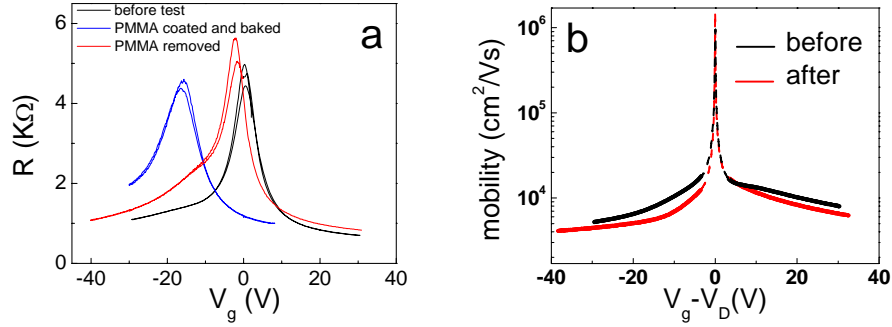

**Figure 2. a**. Gate voltage dependence of the resistivity at different stages of the lithography process, indicated by the legends. b. Gate voltage dependence of mobility before the PMMA coating test and after the PMMA removal.

In order to understand the effect of Aluminum contacts on the Fermi energy in graphene we coated a freshly made graphene device with a partially oxidized aluminum oxide film. We used e-beam evaporation of aluminum in oxygen environment, keeping the graphene sample in a local oxygen pressure of ~1e-4 Torr. The Aluminum was evaporated at a rate of ~0.3 Å/sec, which was mostly not yet fully oxidized when coated onto the graphene device. Aluminum oxide films grown by this method were characterized and controlled to be transparent and insulating below its breakdown field. Yet incompletely oxidized aluminum particles, which are still present in these films, act as dopants. Figure 3b shows the $\rho(V_g)$ dependence of the graphene device before aluminum oxide coating, after the aluminum oxide coating, and after 2 steps of further oxidation of the aluminum oxide film. After the initial aluminum oxide coating we observe a large shift of the maximum in the $\rho(V_g)$ curve to negative gate voltages indicating electron doping. The hysteresis is a result of un-settled charges in the aluminum oxide film. This doping effect can be attributed to incompletely oxidized particles, as confirmed by reduction of the effect upon deterioration of the transport properties of graphene in the presence of the $Si/SiO_2$ substrate is primarily due to trapped charges either within the $SiO_2$ substrate or at the interface between graphene and $SiO_2$. The resulting charge inhomogeneity leads to enhanced long range Coulomb scattering[14] which becomes especially important near the neutrality point where screening is poor. In addition, the atomic roughness of the substrate introduces short range scattering centers and may contribute to quench-condensation of ripples within the graphene layer[15]. To study the effect of the substrate on the quality of graphene, we developed techniques for fabricating suspended graphene (SG) devices with multiple contacts, which allowed transport measurements for sample characterization. The SG devices were fabricated from conventional non-suspended graphene (NSG) devices with Au/Ti leads deposited on $Si/SiO_2$ (300nm) substrates. The freshly prepared NSG device is coated with PMMA followed by an additional e-beam lithography step to open two small windows (typically 0.2 ~ 0.5μm squares) in the PMMA on the two sides of the graphene channel (illustrated in Figure1). The samples were then immersed in 7:1 ($NH_4F$: HF) buffered oxide etch (BOE). Etching was done at $25C^0$ for 6.5 min. Due to weak coupling between graphene and the substrate, capillary action draws the etchant underneath the whole graphene film. Therefore, etching actually starts in the entire graphene channel shortly after the sample is immersed. Because the etching is isotropic it causes the whole device (graphene together with the leads attached to it) to become suspended. When the etching is complete, the etchant is replaced by DI water, followed by hot acetone (to remove the PMMA) and finally hot isopropanol, with the sample remaining in the liquid at all times. Finally, the sample is taken out of the isopropanol and left to dry. At this point one would expect that the very fragile suspended device would be destroyed by wicking of the liquid as it evaporates. However, due to the small surface tension of hot isopropanol, devices with channel length smaller than 1μm were found to survive the process with a high success rate. Figure 4 shows the SEM image of an actual SG device. The SG samples were baked in forming gas (Ar/$H_2$) at $200^0C$ for 1 hour to remove any remaining organic residue and water molecules. In order to avoid contamination this step was promptly followed by transfer to the low pressure measurement cell.

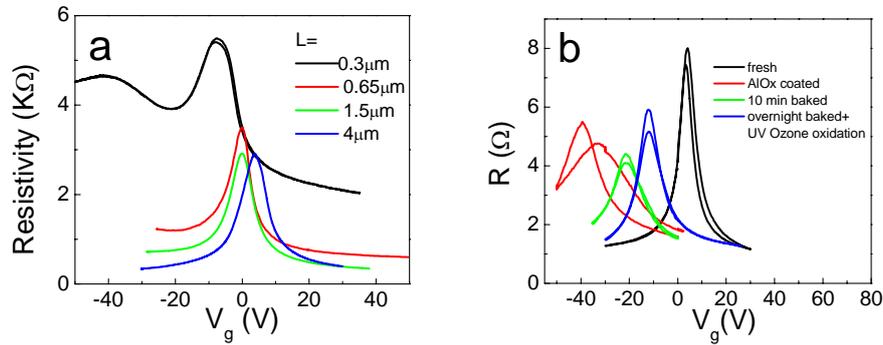

**Figure 3. a**. Gate voltage dependence of resistivity for various samples with different channel length as indicated in the legends. **b.** Resistance as a function of gate voltage for a graphene device. The device was measured before AlOx coating (black). Then it was coated with incompletely oxidized AlOx, baked in oxygen for 10 min (100C), and finally further oxidized by overnight oxygen baking and ozone oxidation. At the end of each step, the sample was measured to characterize its quality.

To characterize the relation between the gate voltage and the induced carrier density, low temperature magneto-transport measurements were carried out. The carrier density at applied gate voltages were related to the Shubnikov-deHaas (ShdH) oscillations by $n = (4e/h)B_F(V_g)$, where $B_F$ is the frequency of the ShdH oscillation. Here we found from the $n(V_g)$ relation that the dielectric constant of the gate dielectric is close to 1 (as is for vacuum), indicating complete removal of the $SiO_2$ underneath the graphene channel.

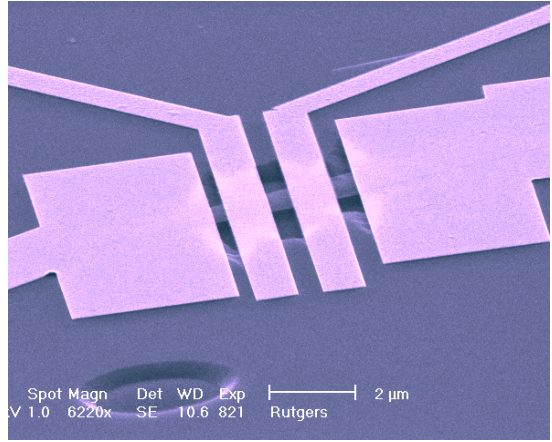

**Figure 4.** SEM image of a suspended graphene device.

To characterize the quality of the SG devices, we focused on the carrier density dependence of the resistivity, $\rho(n)$, in zero magnetic field in the temperature range 4.2K to 250K (Figure 5a). Compared to NSG devices with similar geometry, the SG samples show much sharper carrier density dependence of the resistivity around the DP at low temperatures. At the lowest temperature, 4.2K, the hole branch half width at half maximum (HWHM) is $\delta V_g \sim 0.15$ V, $\delta n \sim 3.2 \ 10^9 cm^{-2}$ for the sample with channel length L = 0.5 μm, is almost one order of magnitude narrower than that of the best NSG samples published so far[4, 16]. This is directly seen (Figure 5b) in the side-by-side comparison of the $\rho(n)$ curves for SG and NSG samples of the same size and taken from the same graphite crystal. The HWHM of the NSG sample shown here is $\delta V_g \sim 3V$ ($\delta n \sim 2.2 \ 10^{11} cm^{-2}$). The sharp carrier density dependence of the resistivity is a direct consequence of the greatly reduced potential fluctuations at the DP. These potential fluctuations induce electron-hole puddles and broaden the gate (carrier density) dependence of the resistivity at the charge neutrality point clearly seen in the NSG device. The reduction of the potential fluctuation and electron-hole puddles also results in the strong temperature dependence of the maximum resistivity at the DP in the SG samples. This is in stark contrast to NSG samples, where the maximum resistivity saturates below $\sim 200K^{17}$, because the residual carrier population in the electron-hole puddles induced by the potential fluctuations is much larger than the thermally activated carriers at these temperatures

A direct consequence of the low level of charge inhomogeneity in the SG samples is that one can follow the intrinsic transport properties of Dirac fermions much closer to the DP than is possible with any NSG samples fabricated to date. In Figure 5c we compare the carrier density dependence of mobility $\mu = \sigma/ne$ for SG and NSG samples to that of the calculated mobility in a ballistic device. For T<100K, at low carrier densities (just outside the puddle regime) the maximum mobility of the SG samples exceeds 100,000 cm$^2$/Vs. compared to ~ 2,000-20,000 cm$^2$/Vs in the best NSG samples. We note that the SG mobility is approaching the calculated value of ideal ballistic devices. Since at low densities the mobility is mostly determined by Coulomb scattering[18] (short range scattering is very weak near the DP due to the small density of states[19]), the difference in mobility between the SG and NSG samples is naturally attributed to substrate-induced charge inhomogeneity. The removal of the substrate in the SG samples eliminates the primary source of Coulomb scattering, the trapped charges. At high carrier densities ($n > 4 \times 10^{11}$ cm$^{-2}$), the mobility in the two types of samples becomes comparable (~10000 cm$^2$/Vs) indicating that short range scattering becomes dominant. The short range scattering can be attributed to imperfections in the graphene layer reflecting defects in the parent graphite crystal or could be introduced during the fabrication process. Both sources of defects can in principle be reduced to produce SG samples with even better quality.

Figure 5c illustrates the density dependence of the mean free path $mfp = \sigma h/2e^2 k_F$ for the SG sample at the indicated temperatures. The negative slope and absence of T dependence for T < 100K, suggest that scattering is predominantly by short range scatterers. For T>100K, the slopes become increasingly positive, suggesting thermally induced long range scattering. Such long range scattering cannot be attributed to charged impurities because such mechanism is expected to be independent of T[17]. Possible explanations include scattering by thermally excited ripples[15] and ripple induced charge inhomogeneity[20]. However, more work is needed to understand the scattering mechanism in this regime.

3. Discussion

We studied the effect of extrinsic factors on the quality of the graphene devices including material quality, lithography process, metallic leads and substrate. This work suggests that material defects in graphite can reduce the mobility and possibly the minimum conductivity. Higher mobility can be achieved by using better quality graphite. The lithographic process leaves a polymer residuewhich may induce Coulomb scattering. However, our tests suggested that the effect of the residue is minor and that it can in principle be eliminated by removing the residue with an appropriate cleaning method, such as baking in forming gas or UHV baking. The contact leads can play an important role and can modify the transport properties of short graphene junctions. Further study needs to be carried out to fully understand this effect and to find the most favorable lead geometry and material.

Our work suggests that the most important limitation on the quality of graphene devices is imposed by trapped charges associated with the substrate. These could be charged scatterers trapped inside the SiO$_2$ dielectric or at the graphene-substrate interface. The former may be reduced by pre-treatment of the substrate (annealing, for example) before the graphene deposition. The latter may be the most important reason for the observed large variations in device quality, because graphene deposition, hence the trapped charged scatterers (from water molecules, organic contamination, etc.) between the deposited graphene and the substrate, vary greatly from device to device. To reduce these effects it is important that the deposition be done in a clean and well controlled manner. It is also important to notice that even though the SG devices show excellent quality at low temperatures, their room temperature characteristics appear to be similar to that of the NSG devices, possibly due to the thermally induced corrugation (ripples) scattering. Therefore, in order to fabricate high quality room temperature graphene devices, it is still necessary to have a substrate in order to suppress the ripples.

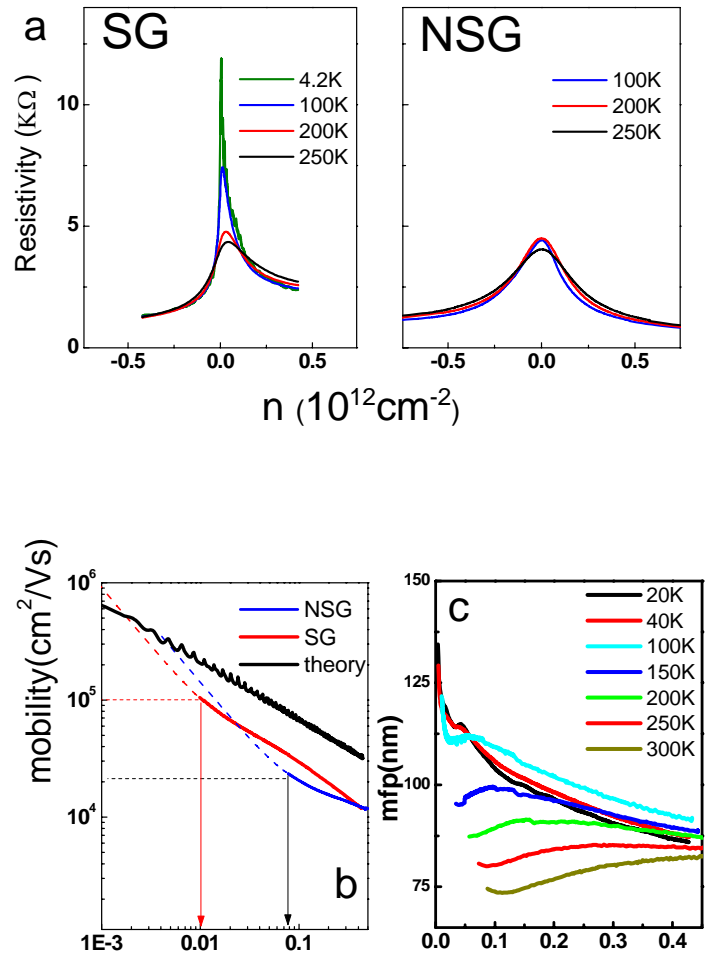

**Figure 5. a** Side-by-side comparison between carrier density dependence of the resistivity at various temperatures for NSG and SG devices with 500nm channel length. **b**. Carrier density dependence of mobility for 500nm NSG and SG at T=100K compared with calculated values for a ballistic graphene device with the same geometry. **c**. Carrier density dependence of mean free path for SG device at various temperatures.

In summary, the work described here indicates a number of necessary conditions that are needed to achieve high quality graphene devices reproducibly. a. High quality single crystal graphite should be used for extracting graphene layers. b. Trapped charges in the gate dielectric have to be eliminated - this could in principle be accomplished by a proper choice a dielectric substrate, but more work is needed to identify such substrtae. c. Trapped impurities at the graphene-substrate interface, such water and organic residue, should be avoided by carrying out the deposition in a controlled and clean environment. We expect that devices fabricated under these guidelines will exhibit many novel physical properties that are expected to emerge when the motion of Dirac fermion is ballistic.

## Acknowledgments

We thank G. Li, Z. Chen for discussions; S.W. Cheong and M. Gershenson for use of AFM and e-beam; V. Kiryukhin for HOPG crystals; A. H. Castro-Neto and F. Guinea for useful discussions. Work supported by DOE, Office of the Army and ICAM.